\documentclass[aps,superscriptaddress,nofootinbib]{revtex4}
\usepackage{amsmath}
\usepackage{graphicx}
\usepackage{color}
\usepackage{slashed}

\begin{document}

\title{Area spectrum of extremal black holes with warped AdS near-horizon geometry}
\author{Wen-Yu Wen}\thanks{%
E-mail: steve.wen@gmail.com}
\affiliation{Department of Physics and Center for High Energy Physics, Chung Yuan Christian University, Chung Li City, Taiwan}
\affiliation{Leung Center for Cosmology and Particle Astrophysics\\
National Taiwan University, Taipei 106, Taiwan}

\begin{abstract}
In this letter, we provide an alternative method to study the area spectrum of certain classes of extremal black holes which have near-horizon geometry as warped AdS.  We argue that previous methods which are based on the existence of quasinormal modes may not be applicable in the extremal limit.  The topology difference of the near-horizon geometry between non-extremal and extremal black holes implies a separate treatment is needed to study the area discreteness in the extremal limit.  To be specific, we will study area spectrum of supersymmetric BMPV black holes/black rings and Reissner-Norstr\"{o}m black holes at the extremal limit.     Inspired by the recently established Kerr/CFT and RN/CFT correspondence, we propose a new way to quantize the area regardless of the (non-)existence of quasinormal modes or zero Hawking temperature.  At last, we propose a dilute gas model and harmonic oscillator model which have same degrees of freedom as the dual CFT.
\end{abstract}

\maketitle




A finite size system often displays a discrete energy spectrum for quantum fluctuation.   It was suggested that since the dynamics of a black hole is uniquely determined by its charge(s), which is closely related to the finite region enclosed by the horizon, one also expects the mass or area spectrum to display similar discreteness\cite{Bekenstein:1995ju,Bekenstein:1997bt}.  There were many proposals to obtain area spectrum for various black holes since then.  In the following, we would like to motivate our approach by first recalling some simple facts on near horizon geometry of extremal Reissner-Norstr\"{o}m (RN) black holes.

The metric of 4D RN black hole reads:
\begin{eqnarray}\label{eq:RNbh}
ds_4^2&=&-f(r)dt^2+\frac{dr^2}{f(r)}+r^2 d\Omega_2^2,\nonumber\\
A &=& -\frac{2Q}{r} dt,\qquad f(r) = 1-\frac{2M}{r}+\frac{Q^2}{r^2}
\end{eqnarray}
where the $M$ and $Q$ are mass and charge parameters in natural units and $G=1$.  In non-extremal limit $M>Q$, the function $f(r)=0$ has two real roots $r_\pm$ and the geometry near outer horizon can be obtained if one sets  $r=r_++\epsilon \frac{\kappa}{2}\rho^2$ and $t=\tau/\epsilon$, then sends $\epsilon \to 0$.  The metric reads,
\begin{eqnarray}\label{eq:nonextremal}
ds^2 &= &-\kappa^2 \rho^2 d\tau^2 + d\rho^2 + r_+^2 d\Omega_2^2,\nonumber\\
\kappa &=& \frac{r_+-r_-}{2 r_+^2},
\end{eqnarray}
which is the Rindler space.  Upon wick rotating $\tau$ and requiring the regularity at origin $\rho=0$, one defines the Hawking temperature as inverse of periodicity $T_H^{-1} =2\pi/ \kappa $.   However, the extremal limit $Q=M$ would bring  (\ref{eq:nonextremal}) into a singular metric thanks to $\kappa=0$.  The discontinuity from non-extremal to extremal limit was recently studied in \cite{Carroll:2009maa}.   In fact, the near horizon geometry of extremal limit can be honestly obtained by starting metric (\ref{eq:RNbh}) with $Q=M$ and setting $r=Q+\epsilon \rho$ and $t=Q^2\tau / \epsilon$.  Then $\epsilon \to 0$ bring the metric into
\begin{equation}\label{eq:extremal}
ds^2=Q^2[-\rho^2d\tau^2 + \frac{d\rho^2}{\rho^2}+d\Omega^2].
\end{equation}
The extremal geometry (\ref{eq:extremal}) is $AdS_2\times S^2$, which has totally different topology from the Rindler space (\ref{eq:nonextremal}), which is $R^2\times S^2$ after wick rotation.  This topological distinction explains that the extremal limit can not be achieved simply by taking $\kappa \to 0$ limit in (\ref{eq:nonextremal}).  We are also amused by the third law of thermodynamics which states that it is impossible to bring a system to absolute zero degree within a finite number of steps.  It might make more sense that at the extremal limit, the entropy of a black hole is obtained by simply counting the available indistinguishable states, if applicable, rather than consulting to other laws of thermodynamics in which finite temperature is involved.  With that being said, the previous method of quantizing horizon area is heavily based on real or imaginary part of quasinormal modes\cite{Hod:1998vk,Hod:2003jn,Barvinsky:2001tw,Maggiore:2007nq,Vagenas:2008yi,
LopezOrtega:2010tg,Setare:2003bd,Setare:2004uu}.  Therefore it seems invalid to apply to a purely quantum system of no thermal dissipation, such as that of extremal limit.  Indeed, some of recent studies do favor the opinion of nonexistence of quasinormal modes in extremal black holes\cite{Chen:2010sn,Crisostomo:2004hj,Hod:2013eea}.  Although recently the application of adiabatic invariant action variable did not use the quasinormal modes or adopt the assumption of small charge\cite{Majhi:2011gz,Zeng:2012wb}, the appearance of Hawking temperature during the integration may still cause it ill-defined if the extremal limit is taken naively.  In the other words, the conventional way of studying extremal black holes via taking near-extremal limit might be too naive or improper if not wrong, though the results might coincidently agree.


In the following, we attempt to obtain the area spectrum of extremal black hole directly from the near horizon geometry (\ref{eq:extremal}) and interpret this discreteness from its holographic dual CFT.  It is known that this extremal geometry can further be uplifted into a warped $AdS_3\times S^2$ metric with $2$-form field as\cite{Garousi:2009zx}
\begin{eqnarray}\label{eq:uplift}
ds_5^2&=&Q^2[-r^2dt^2+\frac{dr^2}{r^2}+ d\Omega_2^2]+ l^2(d\chi+\frac{Q}{l}r dt)^2,\nonumber\\
B_{[2]}&=&-\sqrt{3}l A\wedge d\chi.
\end{eqnarray}
The RN/CFT correspondence states that a chiral CFT can be identified at the asymptotic geometry of uplifted extremal RN metric (\ref{eq:uplift}) and the Bekenstein-Hawking entropy can be reproduced via the Cardy formula\cite{Garousi:2009zx,Chen:2009ht,Chen:2012ps}, in the same spirit of the Kerr/CFT correspondence\cite{Guica:2008mu,Hartman:2008pb}.  



The uplifted extremal RN geometry is a product space which can be viewed as a compact fiber at each location in $AdS_2$.  The topology of fiber is $S^1_l\times S^2_Q$, where $l$ and $Q$ are radii of the circle and sphere.  In particular, each circular fiber at radius $r$ spins in the angular frequency $\omega = Qr/l$ and therefore sets a energy scale $E=\hbar \omega$.  A careful derivation of this energy via adiabatic invariance is present in the appendix.  Since the temperature is absolute zero for the extremal black hole, one expects only quantum fluctuation needs to be considered.  Upon applying the Sommerfeld-Bohr quantization rule for quantum fluctuation on this compact circle, where one requires the circumference is the integer multiple of de Broglie wavelength $\lambda = hc/E$, one obtains the quantization of energy $E=n\hbar c/l$.  For arbitrary but constant $l$, this implies the discretization of radius $r$ in the unit of $\Delta r = \frac{c}{Q}$, which in terms makes the discretization of horizon area (in the unit of $\l_p^2$):
\begin{equation}
\Delta A  = \Delta (4\pi r^2)|_{r=Q} = 8\pi r \Delta r|_{r=Q} = 8\pi Q \Delta r = 8 \pi 
\end{equation}
Following the Bekenstein-Hawking area law, one also obtains the quantization of entropy $S = A/4 = 2 \pi N$, for integer $N$.  The area discreteness $\Delta A=8 \pi$ is same as that of non-extremal RN black holes.  This implies that the change of degrees of freedom is smooth (in integers) from non-extremal to extremal limit.  An extremal RN black hole of charge $Q$ has $N$ quanta of area, where 
\begin{equation}
N = \frac{A}{\Delta A} = \frac{Q^2}{2}.
\end{equation}
Here we would like to examine if $N$ could have similar interpretation in the CFT models. 


The chiral CFT$_2$ dual to the lifted geometry has the left-hand central charge and temperature\cite{Garousi:2009zx}:
\begin{equation}
c_L=\frac{6Q^3}{\l}, \qquad T_L = \frac{\l}{2\pi Q}. 
\end{equation}
The entropy is given by either form of Cardy formula
\begin{equation}
S = \frac{\pi^2}{3}c_L T_L = 2\pi \sqrt{\frac{c_Lh_L}{6}},
\end{equation}
from which one can derive $h_L = \frac{Q\l}{4}$.   One recalls that $h_L$ is related to total energy of ground state $L_0$ and Casimir energy $\frac{c_L}{24}$ by $h_L=L_0-\frac{c_L}{24}$.  One obtains the total energy as 
\begin{equation}
L_0 = \frac{1}{4}(Q\l +\frac{Q^3}{\l}).
\end{equation}
%


If we fix the radius of lifted circle to be $\l = \l_p=1$ as suggested in \cite{Garousi:2009zx}, while regarding $N$ as the level of harmonic oscillator with discrete energy $\hbar\omega = \frac{Q}{2}$, then $L_0$ is exact the energy of the oscillator:
\begin{equation}
L_0 = \hbar\omega (N+\frac{1}{2}).
\end{equation}
%

On the other hand, if we fix the radius of lifted circle as $\l=Q$\cite{Chen:2009ht}, one obtains $L_0=\frac{Q^2}{2}=N$.  In this model, one views the system composed of non-interacting dilute gas of $N$ particles.  Each gas particle carries energy $1$ in nature units according to the principle of equipartition of energy.

So far we have succeeded in quantizing horizon area by applying Sommerfeld-Bohr quantization rule along the lifted circle.  Some comments are in order: at first, the idea of quantizing angular momentum to obtain area spectrum first appeared in the study of non-extremal RN black holes\cite{Ropotenko:2009mh}.   However, the near horizon geometry (\ref{eq:nonextremal}) is essential in their approach, so it cannot be applied to the extremal limit.  Secondly, the energy scale set up by the lifted circle,  $E=\hbar \omega$, is linearly proportional to $r$, which is consistent with the Holographic Principle which states that the holographic direction $r$ is regarded as energy scale in the dual CFT, where small (large) $r$ is associated with IR (UV) limit.  Thirdly, the quantization scheme is independent of the choice of circle size $l$, which is arbitrary in the lifted geometry.   However, with some typical choices of radius, one can have the dual CFT as the dilute gas model at large radius ($Q$) on one hand, and the harmonic oscillator at small radius ($\l_p$) on the other hand.  We suspect that there exists some duality between these two models since they both describe the same degrees of freedom as the $4D$ RN black hole.  In particular, a remnant would have survived after Hawking radiation, if the zero point energy $\frac{Q}{4}$ in the harmonic oscillator model had some physical meaning.

At last, we would like to show that similar quantization scheme could also apply to the Breckenridge-Myers-Peet-Vafa (BMPV) black hole\cite{Breckenridge:1996is} and black ring\cite{Elvang:2004rt}.  As shown in the appendix, the near-horizon geometries of both solutions take the following unified form:
\begin{equation}\label{eq:BMPV}
ds^2=\frac{\l^2}{4} (-r^2dt^2 + \frac{dr^2}{r^2}+d\theta^2+\sin^2\theta d\chi^2)+r_+^2(d\psi^2+\frac{\l}{2r_+}dt)^2,
\end{equation}
where $\frac{\l}{2}$ and $r_+$ are radii of $S^2$ and $S^1$.  The BMPV black hole corresponds to that  two radii are the same.  The metric (\ref{eq:BMPV}) has the same topology as the lifted geometry of extremal RN black hole, and therefore the same quantization treatment is applicable.  Indeed, for arbitrary but fixed $r_+$, the horizon area is quantized in the unit of 
\begin{equation}
\Delta A = 8 \pi^2 r_+,
\end{equation} 
and therefore the number of quanta is given by $N=A/\Delta A = \frac{\l^2}{4}$.  It is known from its dual CFT that the central charge is given by $c=\frac{3\pi\l^3}{2G}$\cite{Isono:2008kx}.  The total energy including the Casimir energy can be calculated to be 
\begin{equation}
L_0 = \frac{2\pi \l}{G}(\frac{\l^2}{4}+\frac{r_+^2}{4}).
\end{equation}
This energy can be regarded as a harmonic oscillator of level $N$ for $\hbar\omega=\frac{2\pi \l}{G}$ when $r_+=\sqrt{2}\l_p=\sqrt{2}$.  At $r_+=\l$, it can also be regarded as a dilute gas of $N$ particles where each particle carries energy $\frac{4\pi\l}{G}$.

A general form of warped AdS geometry occurs in the near-horizon metric of extremal Kerr(-Newman) black holes\cite{Guica:2008mu,Wen:2009qc}:
\begin{equation}
ds^2=\Gamma(\theta)[-r^2dt^2+\frac{dr^2}{r^2}+\alpha(\theta)d\theta^2]+\gamma_1(\theta)(d\phi_1+kr dt)^2+\gamma_2(\theta)d\phi_2^2.
\end{equation}
However, in order to study area spectrum for this extremal metric, it would require a generalized quantization scheme for $\theta$-dependent radii $\Gamma(\theta)\alpha(\theta)$ and $\gamma_i(\theta)$.  We will leave it for future study. 

\appendix
\section{Derivation of energy scale  $E$}
In order to use the adiabatic invariant action variable $H$, we recall the Sommerfeld-Bohr quantization along the lifted circle:
\begin{equation}
nh=\int{Ld\chi}=\int\int_0^L{dL'}d\chi = \int\int_0^H{\frac{dH'}{|\dot{\chi}|}}d\chi,
\end{equation}
for angular momentum $L$ along the circle.  Here we have used the Hamilton equation $\dot{\chi}=\frac{\partial H}{\partial L}$.  For a particle at constant $r$ and $\Omega_2$, its null-like trajectory is 
\begin{equation}
|\dot{\chi}|=\frac{Q}{\l}r
\end{equation}
Then, the above quantization condition becomes 
\begin{equation}
nh = \frac{\l}{Qr}\int_0^H{dH'}\int_0^{2\pi}{d\chi}=\frac{2\pi\l H}{Qr}.
\end{equation}
This implies that the energy is quantized in the unit: 
\begin{equation}
E\equiv \frac{H}{n} = \frac{\hbar Qr}{\l} = \hbar \omega
\end{equation}
as given before.

\section{Near-horizon geometry of the BMPV black holes and black rings}
The near-horizon geometry of the BMPV black holes and black rings have been discussed in \cite{Wen:2009qc}.  Here we present the appendix there for convenience.
\subsection{BMPV black holes}
We start with the BMPV metric\cite{Breckenridge:1996is}:
\begin{eqnarray}
&&ds^2 = -(1+\frac{\l^2}{\rho^2})^{-2}[dt^\prime+\frac{J}{2\rho^2}(d\psi^\prime+\cos{\theta}d\chi)]^2+(1+\frac{\l^2}{\rho^2})(d\rho^2+\rho^2d\Omega_3^2),\nonumber\\
&&d\Omega_3^2=\frac{1}{4}[d\theta^2+\sin^2{\theta}d\chi^2+(d\psi^\prime+\cos{\theta}d\chi)^2].
\end{eqnarray}
The near-horizon geometry is obtained by taking $\rho\to 0$, then 
\begin{equation}
ds^2 = -[\frac{\rho^2}{\l^2}dt^\prime+\frac{J}{2\l^2}(d\psi^\prime+\cos{\theta}d\chi)]^2+\l^2\frac{d\rho^2}{\rho^2}+\l^2d\Omega_3^2.
\end{equation}
To bring it to the desired metric (\ref{eq:BMPV}), one makes the following coordinate transformation:
\begin{eqnarray}
&&r=\rho^2,\qquad  t=\frac{\l^2}{2\sqrt{\l^6-J^2}}t^\prime,\nonumber\\
&&d\psi=d\psi^\prime-\Omega dt,\qquad \Omega \equiv \frac{r^2/\l + Jr^2/\l^4}{\l^2/2-J^2/(2\l^4)},
\end{eqnarray}
and identify $r_+^2=\frac{\l^2}{4}(1-J^2/\l^6)$.

\subsection{Black rings}
We start with the black ring metric\cite{Elvang:2004rt}:
\begin{eqnarray}
&&ds^2 = -f^2(dt+\omega)+f^{-1}[d\rho^2+\rho^2\sin^2{\theta}d\phi^2+\rho^2\cos^2{\theta}d\psi^2],\nonumber\\
&&f^{-1}=1+\frac{Q-q^2}{\Sigma}+\frac{q^2\rho^2}{\Sigma^2},\qquad \Sigma \equiv \sqrt{(\rho^2-R^2)^2+4R^2\rho^2\cos^2{\theta}},
\end{eqnarray}
where 
\begin{eqnarray}
&&\omega = \omega_\phi d\phi + \omega_\psi d\psi,\nonumber\\
&&\omega_\phi = -\frac{q\rho^2\cos^2\theta}{2\Sigma^2}[3Q-q^2(3-\frac{2\rho^2}{\Sigma})],\nonumber\\
&&\omega_\psi = -\frac{6qR^2\rho^2\sin^2{\theta}}{\Sigma(\rho^2+R^2+\Sigma)}-\frac{q\rho^2\sin^2{\theta}}{2\Sigma^2}[3Q-q^2(3-\frac{2\rho^2}{\Sigma})].
\end{eqnarray}
Following the discussion in \cite{Elvang:2004rt,Elvang:2004ds}, one can identify its near-horizon geometry as that of extremal BTZ black hole with mass $M_{BTZ}=2L^2/q^2$ and spin $J_{BTZ}=qM_{BTZ}$ as follows:
\begin{eqnarray}
&&ds^2 = \frac{q^2}{4}\frac{dr^2}{r^2}+\frac{4L}{q}rdtd\psi^\prime + L^2 d\psi^\prime + \frac{q^2}{4}(d\theta^2+\sin^2{\theta}d\chi^2),\nonumber\\
&&L\equiv \sqrt{3[\frac{(Q-q^2)^2}{4q^2}-R^2]},
\end{eqnarray}
which gives the desired metric (\ref{eq:BMPV}) with $q=\l$ and $r_+=L$.

\begin{acknowledgments}
This work is supported in part by the Taiwan's National Science Council (grant No. 102-2112-M-033-003-MY4) and the National Center for Theoretical Science. 
\end{acknowledgments}



\begin{thebibliography}{99}

\bibitem{Bekenstein:1995ju} 
  J.~D.~Bekenstein and V.~F.~Mukhanov,
  ``Spectroscopy of the quantum black hole,''
  Phys.\ Lett.\ B {\bf 360}, 7 (1995)
  [gr-qc/9505012].

\bibitem{Bekenstein:1997bt} 
  J.~D.~Bekenstein,
  ``Quantum black holes as atoms,''
  gr-qc/9710076.

\bibitem{Carroll:2009maa} 
  S.~M.~Carroll, M.~C.~Johnson and L.~Randall,
  ``Extremal limits and black hole entropy,''
  JHEP {\bf 0911}, 109 (2009)
  [arXiv:0901.0931 [hep-th]].

\bibitem{Hod:1998vk} 
  S.~Hod,
  ``Bohr's correspondence principle and the area spectrum of quantum black holes,''
  Phys.\ Rev.\ Lett.\  {\bf 81}, 4293 (1998)
  [gr-qc/9812002].

\bibitem{Hod:2003jn} 
  S.~Hod,
  ``Kerr black hole quasinormal frequencies,''
  Phys.\ Rev.\ D {\bf 67}, 081501 (2003)
  [gr-qc/0301122].


\bibitem{Barvinsky:2001tw} 
  A.~Barvinsky, S.~Das and G.~Kunstatter,
  ``Quantum mechanics of charged black holes,''
  Phys.\ Lett.\ B {\bf 517}, 415 (2001)
  [hep-th/0102061].

\bibitem{Maggiore:2007nq} 
  M.~Maggiore,
  ``The Physical interpretation of the spectrum of black hole quasinormal modes,''
  Phys.\ Rev.\ Lett.\  {\bf 100}, 141301 (2008)
  [arXiv:0711.3145 [gr-qc]].

\bibitem{Vagenas:2008yi} 
  E.~C.~Vagenas,
  ``Area spectrum of rotating black holes via the new interpretation of quasinormal modes,''
  JHEP {\bf 0811}, 073 (2008)
  [arXiv:0804.3264 [gr-qc]].


\bibitem{LopezOrtega:2010tg} 
  A.~Lopez-Ortega,
  ``Area spectrum of the d-dimensional Reissner-Nordstrom black hole in the small charge limit,''
  Class.\ Quant.\ Grav.\  {\bf 28}, 035009 (2011)
  [arXiv:1003.4248 [gr-qc]].

\bibitem{Setare:2003bd} 
  M.~R.~Setare,
  ``Area spectrum of extremal Reissner-Nordstrom black holes from quasinormal modes,''
  Phys.\ Rev.\ D {\bf 69}, 044016 (2004)

  ``Nonrotating BTZ black hole area spectrum from quasinormal modes,''
  Class.\ Quant.\ Grav.\  {\bf 21}, 1453 (2004)

  ``Near extremal Schwarzschild-de Sitter black hole area spectrum from quasinormal modes,''
  Gen.\ Rel.\ Grav.\  {\bf 37}, 1411 (2005)

\bibitem{Setare:2004uu} 
  M.~R.~Setare and E.~C.~Vagenas,
  ``Area spectrum of Kerr and extremal Kerr black holes from quasinormal modes,''
  Mod.\ Phys.\ Lett.\ A {\bf 20}, 1923 (2005)
  [hep-th/0401187].

\bibitem{Chen:2010sn} 
  B.~Chen and J.~-j.~Zhang,
  Phys.\ Lett.\ B {\bf 699}, 204 (2011)
  [arXiv:1012.2219 [hep-th]].



\bibitem{Crisostomo:2004hj} 
  J.~Crisostomo, S.~Lepe and J.~Saavedra,
  Class.\ Quant.\ Grav.\  {\bf 21}, 2801 (2004)
  [hep-th/0402048].

\bibitem{Hod:2013eea} 
  S.~Hod,
  Physics Letters B 713, {\bf 505} (2012)
  [arXiv:1304.6474 [gr-qc]].

\bibitem{Majhi:2011gz} 
  B.~R.~Majhi and E.~C.~Vagenas,
  ``Black hole spectroscopy via adiabatic invariance,''
  Phys.\ Lett.\ B {\bf 701}, 623 (2011)
  [arXiv:1106.2292 [gr-qc]].

\bibitem{Zeng:2012wb} 
  X.~-X.~Zeng and W.~-B.~Liu,
  ``Spectroscopy of a Reissner-Nordstr\'{o}m black hole via an action variable,''
  Eur.\ Phys.\ J.\ C {\bf 72}, 1987 (2012)
  [arXiv:1204.1699 [gr-qc]].

\bibitem{Garousi:2009zx} 
  M.~R.~Garousi and A.~Ghodsi,
  ``The RN/CFT Correspondence,''
  Phys.\ Lett.\ B {\bf 687}, 79 (2010)
  [arXiv:0902.4387 [hep-th]].

\bibitem{Chen:2009ht} 
  C.~-M.~Chen, J.~-R.~Sun and S.~-J.~Zou,
  ``The RN/CFT Correspondence Revisited,''
  JHEP {\bf 1001}, 057 (2010)
  [arXiv:0910.2076 [hep-th]].

\bibitem{Chen:2012ps} 
  B.~Chen and J.~-J.~Zhang,
  ``RN/CFT Correspondence From Thermodynamics,''
  JHEP {\bf 1301}, 155 (2013)
  [arXiv:1212.1959].

\bibitem{Guica:2008mu} 
  M.~Guica, T.~Hartman, W.~Song and A.~Strominger,
  ``The Kerr/CFT Correspondence,''
  Phys.\ Rev.\ D {\bf 80}, 124008 (2009)
  [arXiv:0809.4266 [hep-th]].

\bibitem{Hartman:2008pb} 
  T.~Hartman, K.~Murata, T.~Nishioka and A.~Strominger,
  ``CFT Duals for Extreme Black Holes,''
  JHEP {\bf 0904}, 019 (2009)
  [arXiv:0811.4393 [hep-th]].

\bibitem{Ropotenko:2009mh} 
  K.~Ropotenko,
  ``Quantization of the black hole area as quantization of the angular momentum component,''
  Phys.\ Rev.\ D {\bf 80}, 044022 (2009)
  [arXiv:0906.1949 [gr-qc]].

\bibitem{Breckenridge:1996is} 
  J.~C.~Breckenridge, R.~C.~Myers, A.~W.~Peet and C.~Vafa,
  ``D-branes and spinning black holes,''
  Phys.\ Lett.\ B {\bf 391}, 93 (1997)
  [hep-th/9602065].


\bibitem{Elvang:2004rt} 
  H.~Elvang, R.~Emparan, D.~Mateos and H.~S.~Reall,
  ``A Supersymmetric black ring,''
  Phys.\ Rev.\ Lett.\  {\bf 93}, 211302 (2004)
  [hep-th/0407065].

\bibitem{Isono:2008kx} 
  H.~Isono, T.~-S.~Tai and W.~-Y.~Wen,
  ``Kerr/CFT correspondence and five-dimensional BMPV black holes,''
  Int.\ J.\ Mod.\ Phys.\ A {\bf 24}, 5659 (2009)
  [arXiv:0812.4440 [hep-th]].

\bibitem{Wen:2009qc} 
  W.~-Y.~Wen,
  ``Holographic descriptions of (near-)extremal black holes in five dimensional minimal supergravity,''
  arXiv:0903.4030 [hep-th].

\bibitem{Elvang:2004ds} 
  H.~Elvang, R.~Emparan, D.~Mateos and H.~S.~Reall,
  ``Supersymmetric black rings and three-charge supertubes,''
  Phys.\ Rev.\ D {\bf 71}, 024033 (2005)
  [hep-th/0408120].

\end{thebibliography}
\end{document}